\documentclass[10pt,letterpaper]{article}
\usepackage[top=0.85in,left=0.85in,footskip=0.75in,marginparwidth=2in]{geometry}

\usepackage[utf8]{inputenc}

\usepackage{cite}
\usepackage{amssymb,amsmath}
\usepackage{float}
\usepackage{booktabs}

\usepackage{nameref,hyperref}

\usepackage[right]{lineno}

\usepackage{microtype}
\DisableLigatures[f]{encoding = *, family = * }

\raggedright
\usepackage{changepage}

\usepackage[aboveskip=1pt,labelfont=bf,labelsep=period,singlelinecheck=off]{caption}

\makeatletter
\renewcommand{\@biblabel}[1]{\quad#1.}
\makeatother

\usepackage{lastpage,fancyhdr,graphicx}
\usepackage{epstopdf}
\fancyheadoffset[L]{2.25in}
\fancyfootoffset[L]{2.25in}

\usepackage{color}

\definecolor{Gray}{gray}{.25}

\usepackage{graphicx}

\usepackage{sidecap}

\usepackage{wrapfig}
\usepackage[pscoord]{eso-pic}
\usepackage[fulladjust]{marginnote}
\reversemarginpar

\begin{document}
\vspace*{0.35in}

\begin{flushleft}
{\Large
\textbf\newline{Investigating Data-Driven Systems as Digital Twins: Numerical Behavior of Ho--Kalman Method for~Order~Estimation}
}
\newline
\\
Alexios Papacharalampopoulos\textsuperscript{1,*}
\\
\bigskip
\bf{1} Laboratory for Manufacturing Systems and Automation (LMS), Department of Mechanical Engineering and~Aeronautics, University of Patras, 265 04 Patras, Greece; apapacharal@lms.mech.upatras.gr; Tel.:~+30-2610-910-160
\\
\bigskip
\textcolor{red}{* THIS PAPER HAS BEEN PUBLISHED IN PROCESSES 2020, UNDER THE SAME TITLE}
\end{flushleft}

\section*{Abstract}
System identification has been a major advancement in the evolution of engineering. As it is by default the first step towards a significant set of adaptive control techniques, it is imperative for engineers to apply it in order to practice control. Given that system identification could be useful in creating a digital twin, this work focuses on the initial stage of the procedure by discussing simplistic system order identification. Through specific numerical examples, this study constitutes an investigation on the most ``natural'' method for estimating the order from responses in a convenient and seamless way in time-domain. The method itself, originally proposed by Ho and Kalman and utilizing linear algebra, is an intuitive tool retrieving information out of the data themselves. Finally, with the help of the limitations of the methods, the potential future outlook is discussed, under the prism of forming a digital twin..



\section{Introduction}
Adaptive control has been quite popular over the last fifty years~\cite{signif1,signif2}, with a variety of methodologies available~\cite{signif3}.
As a matter of fact, as early as 1955, the adaptive techniques have been reported to be widely utilized in industry and this can be come across in literature~\cite{review1958}. The comparative advantage, being the lack of the model, has helped in creating huge related literature.
Even nowadays, with Industry 4.0-like movements across the Globe being the main streams of digitalization trends in industry~\cite{iot1,iot2}, the cognitive functionalities of automation (exploiting Cyber--Physical Systems and Internet of Things) have been integrated to a great extent and the use of adaptive control techniques has been spread even more. Also, there have been reported works~\cite{digtwin1}, where the well-established technology of adaptive system identification has been presented as an underlying technology for a digital twin. 
\par Applications of adaptive control can be found literally everywhere. From domestic applications~\cite{app1}, to engineering~\cite{app2} and manufacturing~\cite{GCbook}, it is highly evident that adaptive control is very useful. Indicatively, recently, identification techniques had been used to model a system response originating from Partial Differential Equations~\cite{CloudControl:papa} and attempting to control it in an empirical, yet adaptive way. In the case of a digital twin formation, automatic operation is highly important, so the identification phase is of utmost importance.
\par A brief, yet full, review on the State-of-The-Art on methods of identification techniques---and more specifically on the issue of choosing the order of the model---reveals initially the use of empirical methods such as trial and error~\cite{MathworksTrial}, estimation utilizing the frequency domain~\cite{frequency}, and maximum a posteriori (MAP) method~\cite{MAPmethod}. Works have been done previous on the choice the the most suitable method~\cite{methodReview}. Furthermore, the co-variances matrix and the residual whiteness are two more methods~\cite{covM,covM2} that are often discussed. Moreover, the set of Bayesian information criterion (BIC)/Akaike information criterion (AIC)/generalized information criterion (GIC) methodologies is another set of methods~\cite{BICaicGIC_1,BICaicGIC2_2} highly utilized; in the literature there has also been a practical comparison between residual sum of squares (RSS) and BIC~\cite{BICvsRSS}. It is worth mentioning at this point that the later method(s) implies the integration of the concept of information.
\par What seems to be missing, however, is a numerical illustration on the simplest, intuitive way to extract such information (meaning the order of the system) from data (the responses themselves). To this end, this work attempts to investigate numerically a simple method for the estimation of the system order, in time domain, utilizing the linear dependence between the sampled data. The concept of the rank of the matrix is utilized as the tool to perform this, as originally suggested by Ho and Kalman~\cite{ho1966effective}. The paper is structured as follows: Firstly, an underlying framework is given. Thus, the methodology of creating a digital twin for a manufacturing process through data-driven models is illustrated. Also, the significance of introducing automated order estimation techniques to such digital twins is pointed out. Next, the Ho--Kalman algorithm is presented and is compared against other methods. In continuation, numerical examples are given on the efficiency of the Ho--Kalman algorithm in various cases. Finally, conclusions are extracted on the significance and the usability of the algorithm.
 \section{Framework}
Regarding manufacturing processes digital twins, it is extremely useful that they are ``near-real-time''~\cite{controlcentric}. This could be defined as having a running time of at least one order of magnitude smaller than process time constants. This way, control and optimization would be feasible. Data-driven problems in particular are very flexible towards this end, as they can be based on adaptive control techniques. However, the estimation of system order may be yet another loophole as proved with numerical examples herein. The framework implied herein is based on such technologies and the order estimation is discussed.
\par
\begin{figure}[H]
\centering
\includegraphics[width=6in]{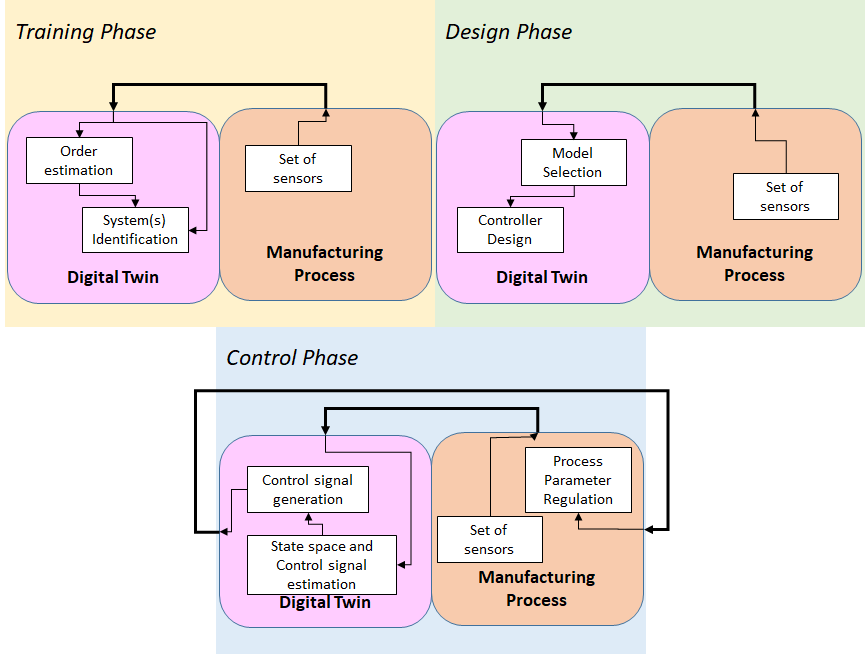}
 \caption{Underlying framework that takes order estimation into account.}
\label{framework}
\end{figure}

  Under the current framework, the Ho--Kalman estimation is considered as an order estimation algorithm. As proved hereafter, the algorithm could be applied with great success in some cases and the scope of the current work is to examine the applicability of this method. The framework, in full description, is described in the list below, given the fact that the main module of the digital twin is a dynamic system. The main functionality is the control of the physical system (process), but other scopes can be defined on top of that, such as running the simulation to respond to What-if scenarios and be able to find proper working conditions (process parameters). Figure~\ref{framework} is used to illustrate the operation of such a control-based digital twin.
\par Training phase:
\begin{enumerate}
  \item Data are aggregated for various cases (i.e., different materials)
  \item Ho--Kalman algorithm is applied to estimate the order of the system
  \item Plain estimation techniques are applied (i.e., mean least squares) to retrieve the transfer function(s)
\end{enumerate}

\par Design phase:
\begin{enumerate}
  \item Sensors are used to detect the model that should be applied
  \item The controller is designed (i.e., Proportional--Integral--Derivative)
\end{enumerate}

\par Control phase:
\begin{enumerate}
  \item Sensors are used to measure input and output of the system
  \item (optional) An observer is used to estimate the state (inner variables) of the system 
  \item The control signal is generated and control is applied (these may be two different steps depending on the implementation)
\end{enumerate}

\section{Method}

As briefly aforementioned, the method investigated here is based on the fact that linear systems response values at time $n$ (in the case of discrete systems) are linear combinations of previous values at time $n-k$, for some $n, k\in\mathbb{N}$. Therefore, the concept of linear independence is exploited, through the concept of ranks of matrices. To achieve this, a matrix is formed, containing translated versions of the response, as shown in Equation (\ref{eq1}), given a response $y[n]$.

\begin{equation}
\tilde{Y}^{N \times N}=
\begin{bmatrix}
    y[0] & y[1] & \dots & y[N-1]  \\
    y[1] & y[2] & \dots & y[N]  \\
 \dots & \dots & \dots & \dots \\
    y[N-1] & y[N]  & \dots & y[2N-2]  
\end{bmatrix}
\label{eq1}
\end{equation}

The order of the system $S$ that had $y[n]$ as a response, is expected to be equal to the rank of this matrix, namely $\rho^{N \times N}=\rho(\tilde{Y}^{N \times N})$. Even in the marginal case where the $N$ is taken to be equal to $M+1$ (with $M$ being the order of the system) it is evident that the rank of the matrix is equal to the order of the system, as shown in Equation (\ref{eq2}).

\begin{equation}
\begin{split}
\tilde{\Phi}^{N \times N}&=
\begin{bmatrix}
    y[0] & \dots & y[N-2] & \sum_{n=0}^{N-2}a_ny[n]  \\
    y[1] & \dots & y[N-1] & \sum_{n=1}^{N-1}a_ny[n]  \\
 \dots & \dots & \dots & \dots \\
    y[N-1] & \dots & y[2N-3] & \sum_{n=N-1}^{2N-3}a_ny[n]  
\end{bmatrix}
\\
\rho(\tilde{\Phi}^{N \times N})&=N-1=M
\end{split}
\label{eq2}
\end{equation}

In the next sections, the numerical performance of this algorithm is investigated with respect to the complexity of the system; the order itself, the system structure and potential noise interfering.

\subsection{Comparison to other methods and Correlation to information}

\begin{figure}[H]
\centering
\includegraphics[width=6in]{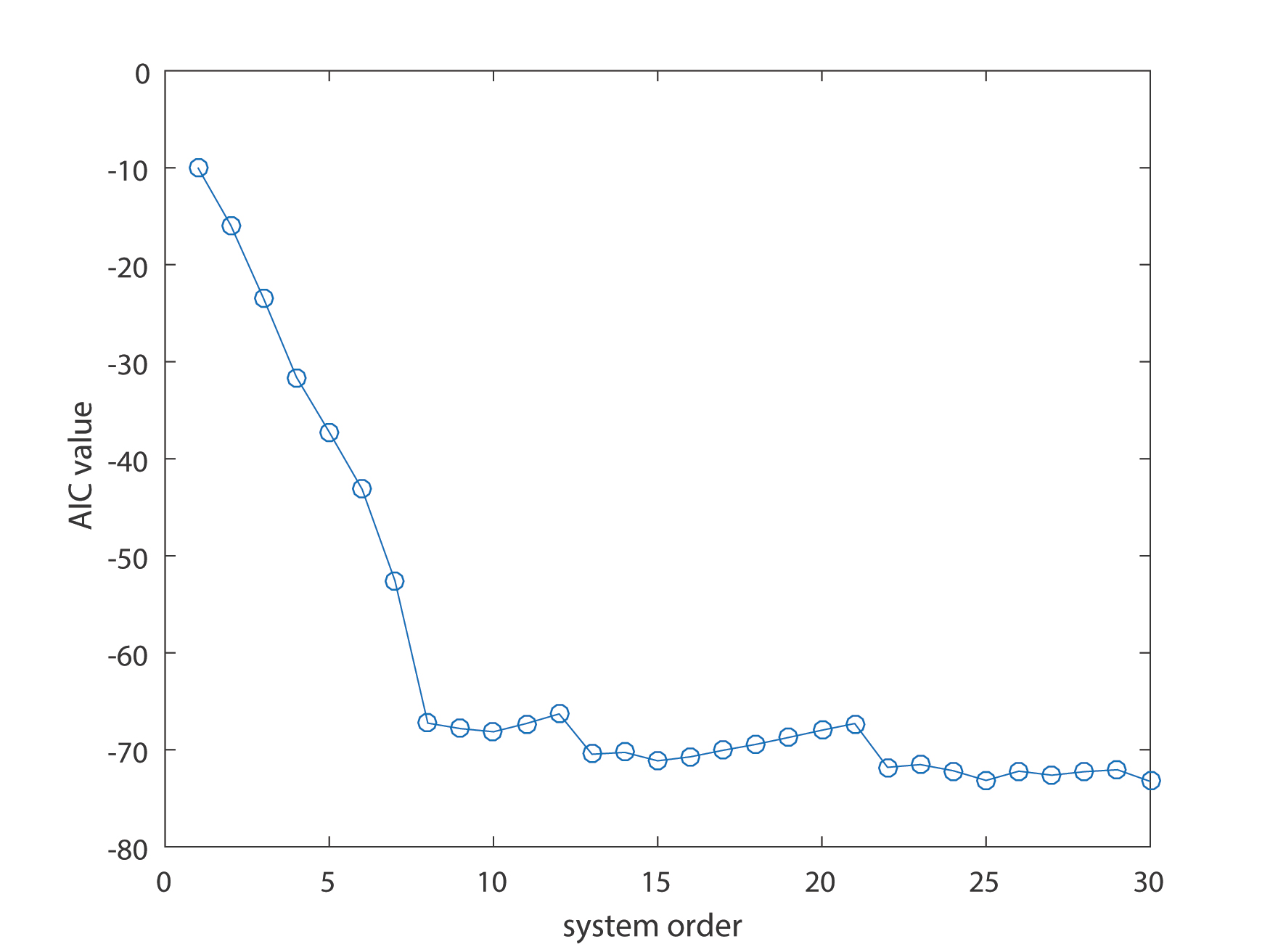}
 \caption{Akaike information criterion (AIC) values as a function of the system order (for system of Equation (3)).}
\label{aic}
\end{figure}

For reasons of completeness, this simple method should be compared against other ones. So, to this end, the following response of Equation (\ref{eq3}) is utilized. The investigated method gives out explicitly (and correctly) an order of 5. However, AIC-based order estimation gives out 8, as shown in Figure~\ref{aic}, while Co-variant Matrix Method leads to inclusive results, as showed in Table \ref{table_cov} (a potential adoption of order 3 could take place).

\begin{equation}
 y_5[n]=\frac{1}{7}\sum_{k=1}^{7}(-1)^{k+1}sin(2 \pi k/3)e^{-\frac{n}{10k}}
 \label{eq3}
\end{equation}

\begin{figure}[H]
\centering
\includegraphics[width=5.5in]{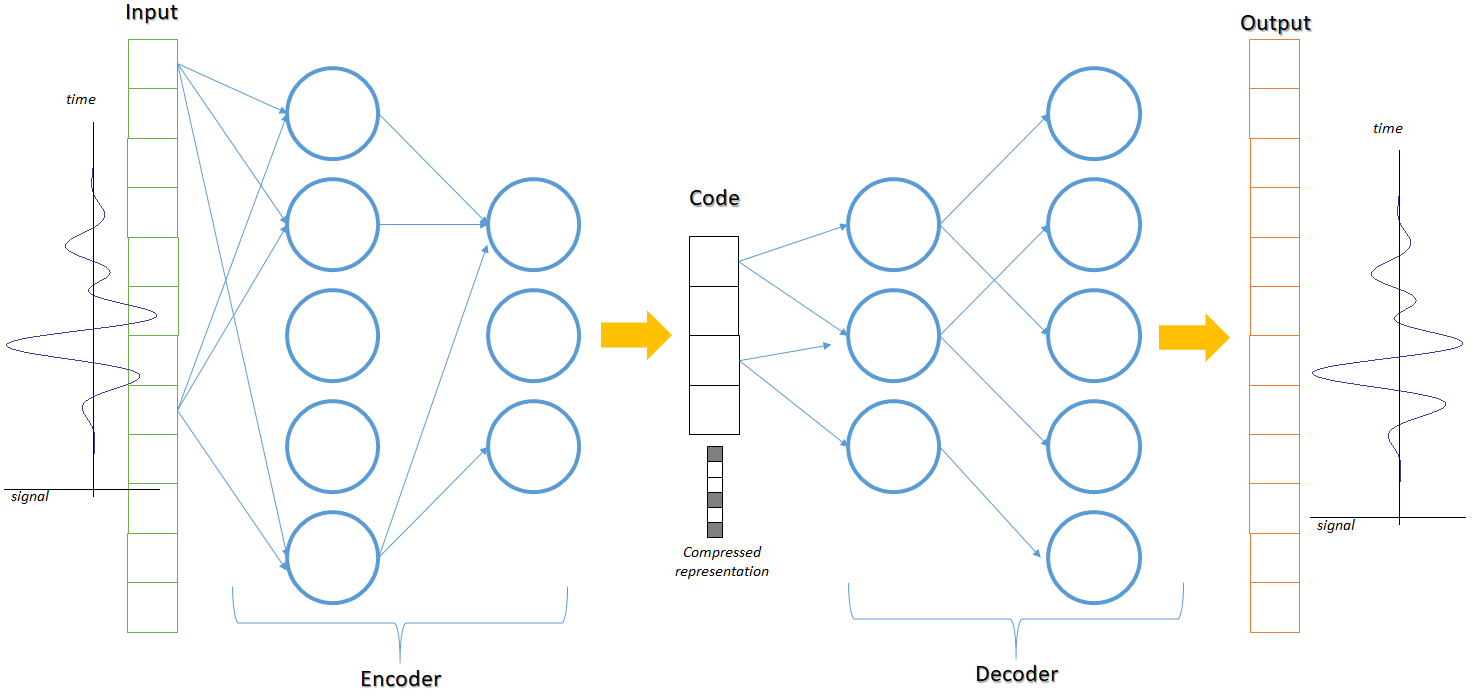}
 \caption{The use of an auto-encoder for data reduction and information compression.}
\label{autoenc}
\end{figure}

\begin{table}[H]
\caption{Co-variance Matrix Results.}
\label{table_cov}
\centering
\begin{tabular}{cc}
\toprule
\textbf{Order of System} & \textbf{Determinant of Co-variance Matrix} \\ 
\midrule
$2$ & $0.00106278$  \\
 $3$ & $1.96022 \times 10^{-12}$  \\
 $4$ & $-5.85109 \times 10^{-26}$  \\
 $5$ & $-4.11625 \times 10^{-40}$ \\
 $6$ & $6.85315 \times 10^{-55}$ \\
 $7$ & $1.71849 \times 10^{-68}$ \\
 $8$ &  $1.78185 \times 10^{-82}$ \\  
\bottomrule
\end{tabular}
\end{table}
This small numerical example has pointed out the numerical superiority of this algorithm---in a case where the method is applicable in its current form. Interestingly enough, the whole point of modelling with a differences equation is of course to be able to reproduce a sequence by a finite (smaller) number of numbers. This slightly reminds one of Chaitin's work~\cite{Chaitin} on linking compression with theory (the concept of statistical inference is also relevant). The only difference herein is that instead of utilizing bits, one tries to compress numbers into numbers, regardless of digits. The same principle lies behind the use of auto-encoders, as illustrated in Figure~\ref{autoenc}. The objective is to utilize a representative feature or equivalently representative compressed signal or image. Thus, the complexity and the dimensionality of the data is reduced. This is very useful for cases where the computational effort has to be reduced. The rank of the responses matrix (even in its full infinite version) is an index of such a complexity (information). So, transformation metrics related to invertibility can also be used, such as the determinant or the eigenvalues distribution. Alternatively, the Lagrangian (or instead a custom Liapunov function) of the system can be used as a different metric. Such a function is of degree higher than linear, thus there is link to correlation matrix method as well. This kind of compression has been very useful in cases where the complexity is simply measured by the rank of the system, such as the case of tool-wear~\cite{indirectTW}.

\section{Numerical Behaviour and Applicability}
So, in the context of finding the numerical limitations of this simple method, various systems have been studied in terms of system order identification. In this section particularly, the applicability and the limitations of the method are shown and discussed through specific paradigms. 

\subsection{Simple Numerical Examples}

To begin with, a first-order system---that would give a response of the form $y_1[n]=Be^{-Qn}$---is utilized (Equation (\ref{5x5})). Thus, a responses matrix of dimensions $5 \times 5$ would be given by $\tilde{Y_1}^{5 \times 5}$.

\begin{equation}
\tilde{Y_1}^{5 \times 5}=B
\begin{bmatrix}
 1 & e^{-Q} &  e^{-2 Q} &  e^{-3 Q} &  e^{-4 Q} \\
  e^{-Q} &  e^{-2 Q} &  e^{-3 Q} &  e^{-4 Q} &  e^{-5 Q} \\
 \dots & \dots & \dots & \dots & \dots \\
  e^{-4 Q} &  e^{-5 Q} &  e^{-6 Q} &  e^{-7 Q} &  e^{-8 Q}
\end{bmatrix}
\label{5x5}
\end{equation}

\begin{figure}[H]
\centering
\includegraphics[width=6in]{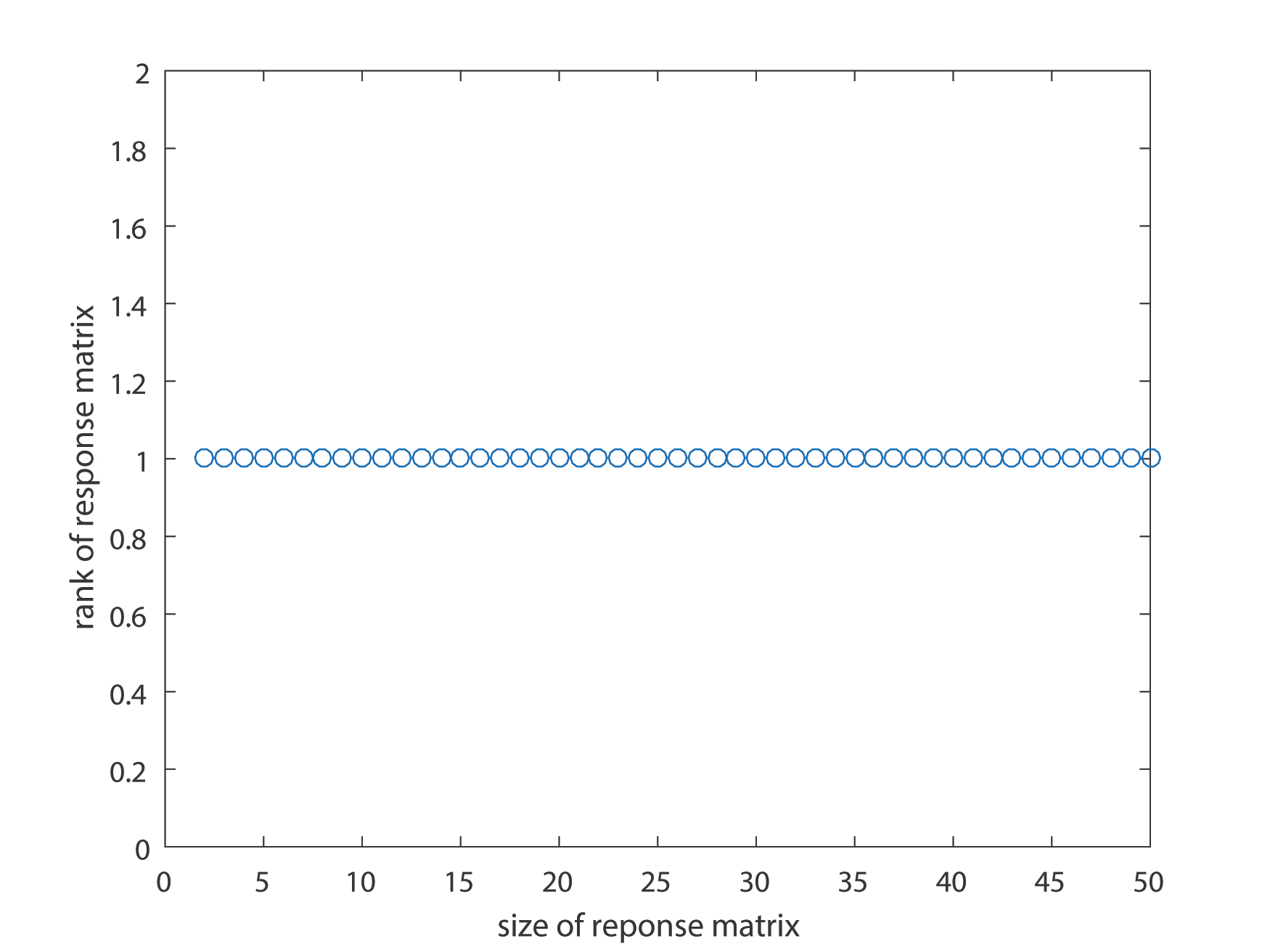}
 \caption{The rank of the responses matrices as a function of the responses matrix dimensions for the case of a first order system.}
\label{fig_sim1}
\end{figure}

Even using symbolic matrices, without specific values, the rank of the matrix, for various dimensions, is equal to 1, as also computationally shown in Figure~\ref{fig_sim1}, for a specific value of $Q$. This is easily proved, as each row (or column) is the product of the previous one with $e^{-Q}$.

\begin{figure}[H]
\centering
\includegraphics[width=6in]{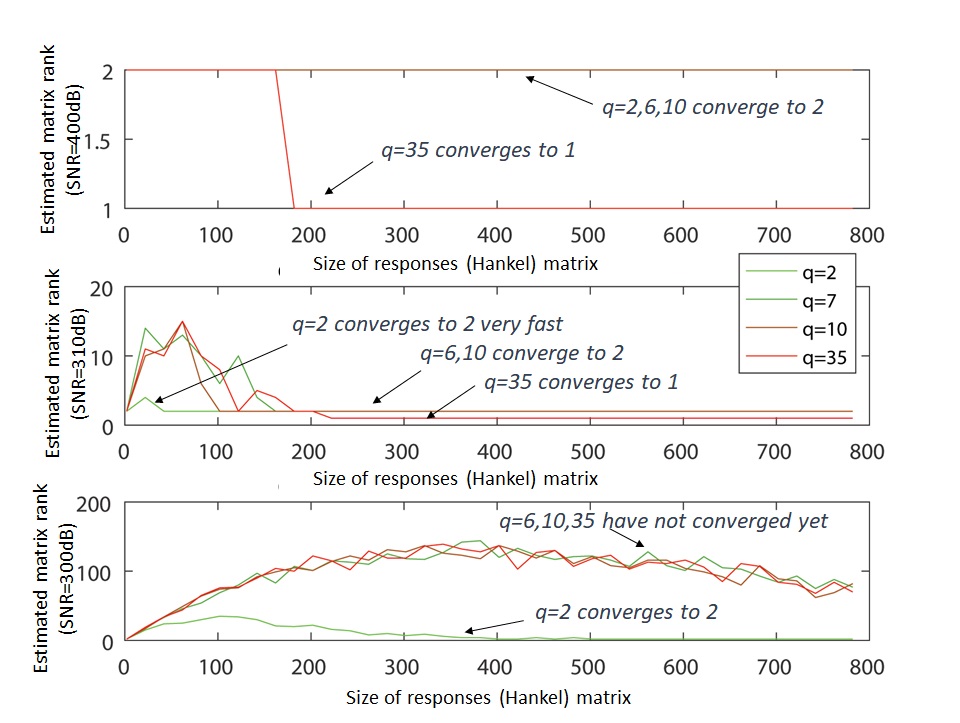}
 \caption{The rank of the responses matrices as a function of the responses matrix dimensions and the proximity of the poles in the case of a second order system. Noise is present.}
\label{fig_sim2}
\end{figure}

So far, everything seems to work well. However, in reality, the sampled values of the responses contain noise, either from measurement, or from sampling itself. Therefore, in this section, noise is going to be regarded, as this is the case in all measured responses. To simulate this, a uniform random number is added after sampling the response, which is regarded in continuous time. Supposing that the continuous-time response is a simplified version of the above one, sampling is applied. In the case where the amplitude of the noise is relatively small, then, as shown in Figure~\ref{fig_sim2}, the convergence is rather rapid. However, if the noise amplitude is increased by one order of magnitude (same Figure), then the convergence becomes much slower. Oddly enough, the unitary signal ($\hat f(t)=1$) has been added to the response on purpose. It has been observed that if the mean value of the response is increased by an offset, then the method converges much faster. Also, the adoption of a row-echelon form of the responses matrix also seems to accelerate the convergence of the method. Furthermore, to study the effect of the poles' proximity on a second-order system, the following response of Equation~(\ref{dpq}) is regarded given that $\delta p=2^{-q}$.

\begin{equation}
y_2[n]=0.5e^{-n/p}+(0.5+\delta p)e^{-n/(p+\delta p)}
\label{dpq}
\end{equation}

The elaboration of such a system has as a goal to study the numerical limitations of the method, as the system tends to be a double-pole system in the limit of $q$ approaching infinity. Simultaneously, the effect of the dominance of one pole is studied. The results are shown in Figure~\ref{fig_sim2}. Evidently, the rank remains equal to $2$, for values of $q<q_0$ and $N\in\{2,..,8\}$, depending also on Signal-to-Noise Ratio (SNR) value. As proved with this numerical study, it seems that for extreme cases of poles' proximity, if the SNR becomes substantially small, then it requires a lot of data for the order estimation alone. For precise calculations in some applications, specifically where this almost double pole affects the controller design, then the response size has to reach up to thousands of samples. Regarding the implementation of the training phase, this affects the use of higher storage capacity, higher memory for processing and faster processors, potentially prohibiting the use of low-performance embedded systems.

\subsection{Performance on Systems of higher order}
To move on to higher order systems, the (arbitrarily chosen) following response (Equation (\ref{f0sinexp})) consisting of $N_0$ terms is considered:

\begin{equation}
 y_N[n]=\frac{1}{N_0}\sum_{k=1}^{N_0}f_0(-n/s_k)
 \label{f0sinexp}
\end{equation}

\begin{figure}[H]
\centering
\includegraphics[width=5in]{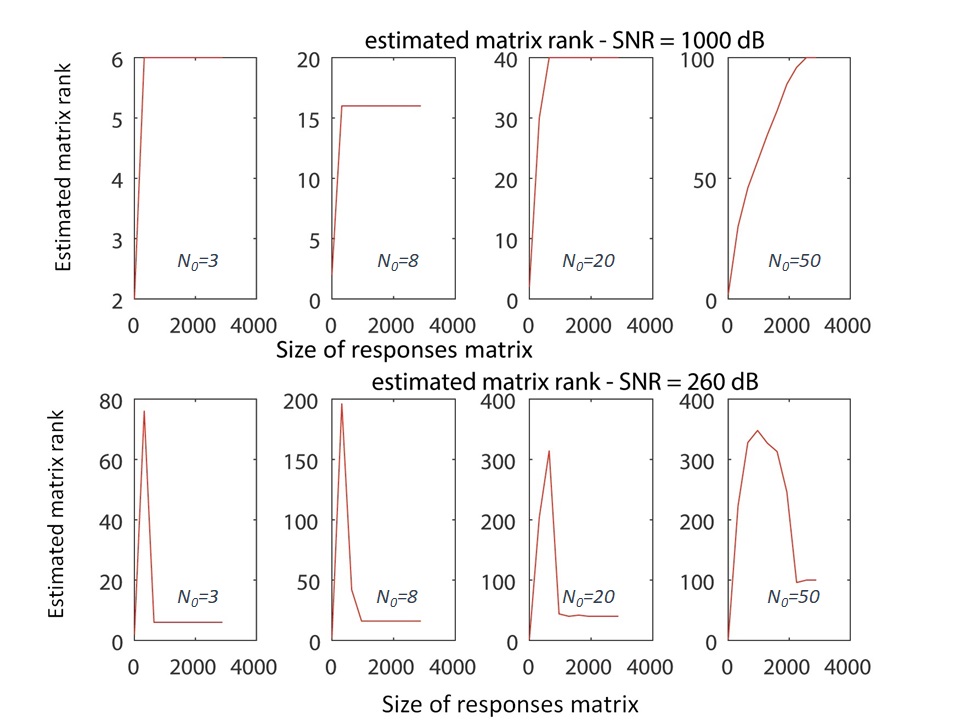}
 \caption{The rank of the responses matrices as a function of the responses matrix dimensions, for the case of Equation 6, where $f_0[n]=sin[n]$ and the upper limit of the sum $N_0$ varies. The values of $N_0$ are shown within the plots.}
\label{high_order_noisy_sinusoidal}
\end{figure}

\begin{figure}[H]
\centering
\includegraphics[width=5in]{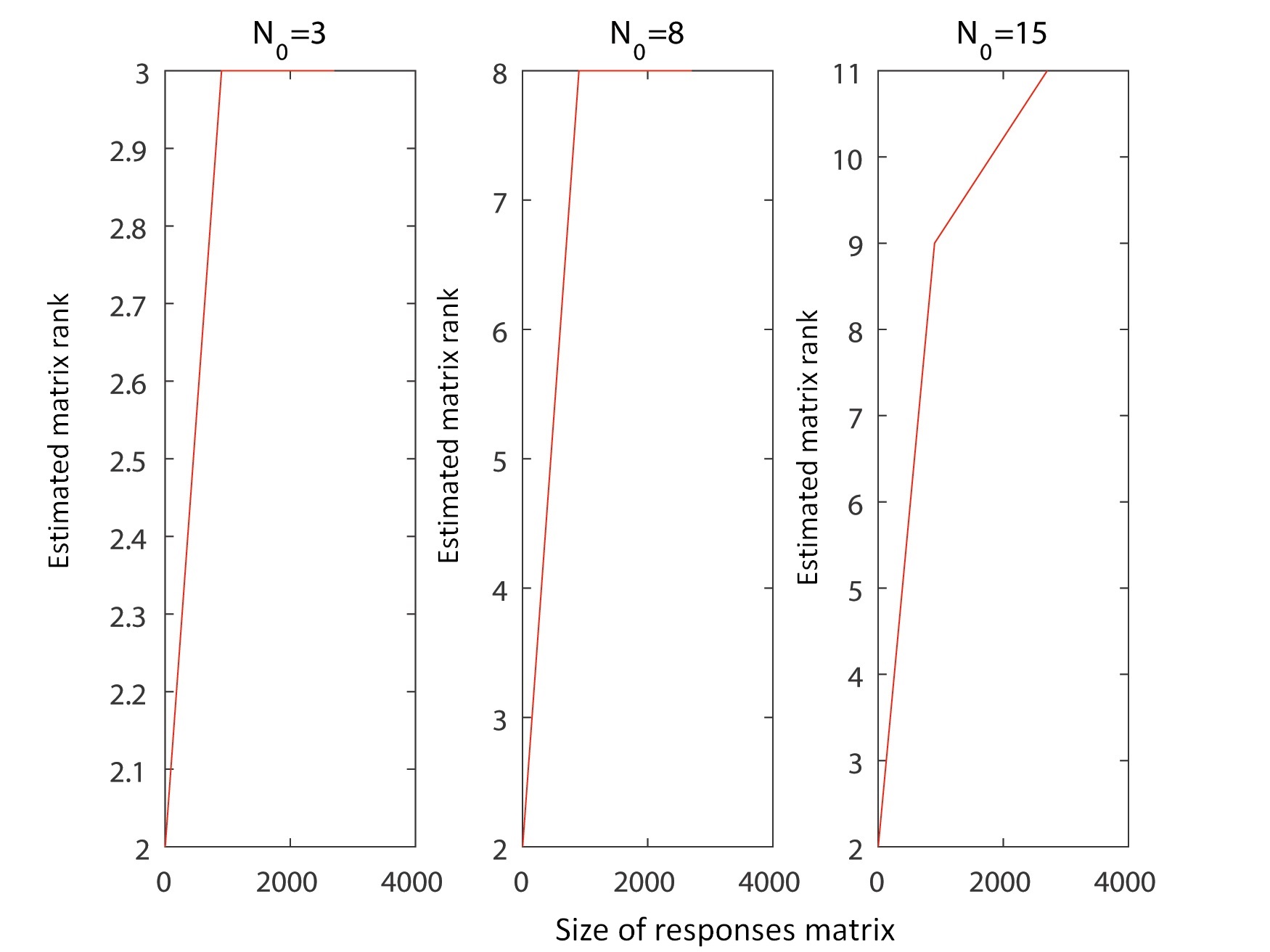}
 \caption{The rank of the responses matrices as a function of the responses matrix dimensions, for the case of Equation 6, where $f_0[n]=e^{-n}$ and the upper limit of the sum $N_0$ varies. The values of $N_0$ are shown above the plots.}
\label{high_order_noisy_exponential}
\end{figure}

The choice of this series as a system leads to a very interesting diagram of $\rho^{N \times N}$ as a function of $N$. The results are given in Figures~\ref{high_order_noisy_sinusoidal} and \ref{high_order_noisy_exponential}, for $f_0(t)=sint$ and  $f_0(t)=e^{-t}$, respectively. It is quite interesting that in the case of sinusoidal functions, some sort of numerical effect takes place. This drives the rank estimation evolution (since it is a function of responses matrix size) to converge to the value of $2N_0$ at a ``faster'' rate. This should be investigated to a further extent, through the consideration of a case of a system of even larger order. This is not the case when dealing with exponential functions, probably due to bad condition number of the responses matrix. Also, since often there can be responses of high order~\cite{toolwear}, i.e., close to 100~\cite{bigorderref}, similar case have also been included here.

\subsection{Non-homogeneous Systems}

Moving on to a different kind of complexity, one can form a matrix for the case of for a non-homogeneous system, such as in the case of $y'(t)+0.9y(t)=e^{-t/8}$. The rank of a $10 \times 10$ responses matrix would be equal to $P+Z=2$, where the $Z$ is the number of Zeros and $P$ the number of poles (1 and 1 respectively). If one augments this matrix to be $11 \times 10$ or $10 \times 11$, padding with (translated) excitation function values to the bottom or to the right, as shown in Equation (\ref{addU}) below, then the rank remains equal to 2. This indicates that the order of the differential equation is equal to $P$ (equal to 1 in this case), as the input and the output have linearly dependent terms.

\begin{equation}
\hat Y^{N \times N}_{aug}=
\begin{bmatrix}
    y[0] & y[1] & \dots & y[N] \\
 \dots & \dots & \dots & \dots \\
    y[N-1] & y[N] & \dots & y[2N-1] \\
    u[0] & u[1] & \dots & u[N] 
\end{bmatrix}
\label{addU}
\end{equation}


\section{Summary and Future Outlook}

The simplest and most intuitive method for estimating the order of the model by the responses themselves in time-domain, utilizing the rank of the responses matrix, has been reviewed as a candidate method for forming automatically a digital twin. It has been proved to be quite successful in many cases, with the efficiency of the algorithm depending highly on the order of the problem as well as the available response dataset size. More specifically, the algorithm appears to work successfully under any kind of system, even though a large amount of data is needed when SNR becomes smaller. The use of this large amount of data seems to cancel the uncertainty introduced by noise. However, its extension towards practical rank estimation algorithms or iterative decomposition of responses is required so that it is able to handle highly noisy data. This  iterative method, may be loosely correlated to the Gram--Schmidt procedure~\cite{GramS}, however, one must have in mind the issue of orthogonality of signals. Nevertheless, the Ho--Kalman algorithm has been proved a very promising tool that may be really useful in slightly noisy data of high dimensionality. In addition to the above, it can be very useful in terms of clustering, in a two-fold way; speeding up computations and offering intuition to a human operator. Also, it seems to be a powerful tool towards compression of data. It has the capability to represent a whole signal through some sort of complexity abstraction.
\par The extensibility of the Ho--Kalman method in non-linear systems ought to be further investigated. The first barrier in this direction is expected to be the very formation of a responses matrix. However, more elaborate tools can be used towards this. Moreover, it seems that as far as the formation of a digital twin is concerned, the use of such a method is promising, however, in the presence of noise, the method has to be combined with the use of another method. Digital twins, either for continuous manufacturing processes, such as plastic extrusion and chemical reactions, or discrete manufacturing processes, such as laser welding and additive manufacturing, will highly benefit from order estimation techniques. This benefit is highly linked to the automated decision making procedure, as the order estimation will not be some kind of fuzzy process that the engineer has to go through. The digital twin will automatically select the values of the (data-driven) model parameters and the control signals generation will be performed automatically and seamlessly.

\vspace{6pt} 





\section*{Acknowledgement}
This work is under the framework of EU Project STREAM-0d. This project has received funding from the European Union’s Horizon 2020 research and innovation program under grant agreement No 723082. The dissemination of results herein reflects only the authors’ view and the Commission is not responsible for any use that may be made of the information it contains.

\begin{figure}[h]
\centering
\includegraphics[width=0.8in]{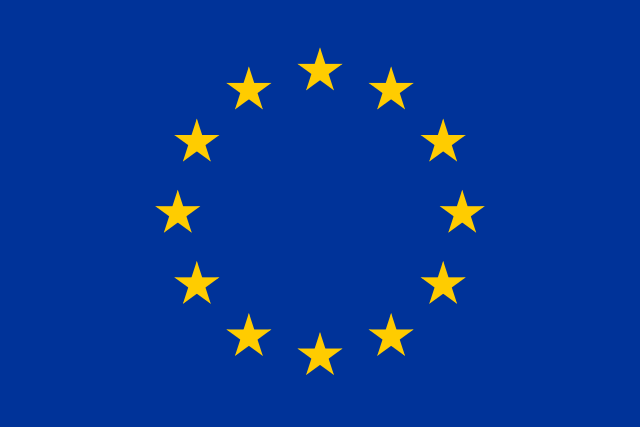}
\label{euflag}
\end{figure}


%





\end{document}